\documentstyle[12pt]{article}
\newcommand{\bi}{\bibitem}
\newcommand{\be}{\begin{equation}}
\newcommand{\ee}{\end{equation}}
\newcommand{\ba}{\begin{eqnarray}}
\newcommand{\ea}{\end{eqnarray}}
\newcommand{\lbl}[1]{\label{eq:#1}}
\newcommand{\ceq}[1]{(\ref{eq:#1})}
\begin{document}

\thispagestyle{empty}
\subsection*{\centerline{ RUSSIAN ACADEMY OF SCIENCES  }}
\subsection*{ \centerline{PETERSBURG NUCLEAR PHYSICS INSTITUTE}}
\hspace*{47mm} ---------------------------------------
\vspace{12mm}

\begin{flushright}
PNPI preprint\\
{\bf TH-15-1997-2158}\\
hep-ph/9703394
\end{flushright}

\vspace{12mm}

\hspace*{27mm} {\large{\bf A.A.Andrianov\footnote[1]{
Department of Theoretical Physics,\,Institute of Physics,\,
University of Sankt-Petersburg,\,
198904,\,Sankt-Petersburg,\,Russia}, \hspace{15mm}  N.V.Romanenko}}
\vspace{5mm}
\nopagebreak[4]
 \begin{center}
 \hspace*{8mm} {\Large\bf Cancellation of Leading Divergencies
 in Left-Right Electroweak Model and Heavy Particles}
\end{center}

\vspace{55mm}
{\Large  \hspace*{50mm} {\sf St.-Petersburg}

\vspace{7mm}
\hspace*{57mm}         {\sf1997 }}
\vspace{7mm}
\pagebreak

\thispagestyle{empty}
\vspace{5mm}
\large
\begin{abstract}
{\sc The fine-tuning principles are analyzed in search for
estimations of heavy particle masses in the left-right (LR)
symmetric model. The modification of
Veltman condition based on the hypothesis of the
compensation between fermion and boson vacuum energies within the LR
Model multiplets is proposed. The hypothesis
 is supplied with the requirement of the stability
under rescaling.
With regard to these requirements
the necessity of existence of right-handed Majorana
neutrinos with masses of order of right-handed gauge bosons is shown
and estimations on the top-quark mass which
are in a good agreement with the experimental value are obtained.}
\end{abstract}

\subsection*{1. Introduction}

  The model with the Left-Right gauge symmetry
$SU(3)_c \otimes SU(2)_L \otimes SU(2)_R \otimes U(1)_{B-L}$  was
proposed  \cite{01,02} as a possible candidate for
the generalization of the Standard Model.  It is rather
attractive both for theoretical
and experimental reasons since it treats left and right
chiral fermions on equal footing at high energies   and the
$P$-parity in this theory is broken spontaneously.
The  possible physical signatures for right-handed
currents can be estimated in high-precision low-energy
experiments such as $\beta$--decay, muon decay, etc.
\cite{2,3}. As well the search for right-handed interactions
in deviations of high-energy experimental data from the
SM predictions remains  an important part of future
collider programmes \cite{3}.

This model can be
considered as a result of $SO(10)$-GUT symmetry breaking.
Its embedding  into a grand
unified scheme \cite{1} can be implemented consistently when the
scale of the discrete symmetry breaking is taken much more than
several TeV.

  In the present paper we would like to apply the
  phenomenological  principles
of the Fine -Tuning (FT) to the  Left-Right (LR) Model  in
order to estimate the heavy mass spectrum of the  theory.
 These principles are based on the
assumption that the theory is actually an effective one
applicable for low energies. Let us formulate them:
\begin{itemize}
\item The strong fine tuning for the scalar field parameters
(v.e.v. and their masses) consists in the
cancellation of large radiative contributions quad\-ratic in ultraviolet
scales which bound the particle spectra in the effective theory
(Veltman-type conditions -\cite{10.1,13.6,13.7,13.8}).
\item The
strong fine tuning for vacuum energies \cite{13.6,13.7,13.8}
deals with the cancellation of large divergencies quartic in
ultraviolet scales which might affect drastically the formation
of the cosmological constant.
\item The RG stability of the
cancellation mechanism under change of ultraviolet scale of
effective theory is provided by the weak fine tuning \cite{13.6,13.7,13.8}.
\end{itemize}
Below on we examine the compatibility of these principles
for the Left-Right
Symmetric model.
Their natural motivation for the LR Model
can be found in  a
more fundamental underlying theory which is free
from nonlogarithmic divergences (for example, in the SUSY
extension of the LR Model).

It will be shown that the FT principles lead to reasonable
values for $m_t$ for a wide range of scales $\Lambda$ and
require the existence of the right-handed Majorana neutrinos
with masses of order of right-handed gauge boson masses
\cite{13.8}.

\subsection*{2. Particle Content of the Theory.}

 We consider LR Model with the
 $SU(3)_c \otimes SU(2)_L \otimes SU(2)_R \otimes U(1)_{B-L}$ gauge group.
 The elecromagnetic charge is expressed in terms of quantum
 numbers of this group by the generalized
 Gell-Mann--Nishijima  formula:  $$Q=T_{3L}+T_{3R}+\frac{Y}{2}$$
 This theory contains three generations of  Standard
 Model fermions with the necessary addition of right-handed
neutrinos.  Fermion assignments for $(T_L,T_R,Y)$ are as
follows:
\be \left[ \begin{array}{c} u \nonumber \\ d
\nonumber \end{array} \right]_{iL}=(\frac{1}{2}, 0,
\frac{1}{3});\; \left[ \begin{array}{c} u \nonumber \\ d
\nonumber \end{array} \right]_{iR}=(0, \frac{1}{2},
\frac{1}{3}) \ee \be \left[ \begin{array}{c} \nu \nonumber \\ l
\nonumber \end{array} \right]_{iL}=(\frac{1}{2}, 0, -1);\;
\left[ \begin{array}{c}
\nu \nonumber \\
l \nonumber \end{array} \right]_{iR}=(0, \frac{1}{2}, -1)
\ee
The gauge sector differs from the SM due to presence of
right-handed gauge bosons. Besides, the Higgs sector of the
model contains more particles than in the SM \cite{3}.

In order to generate fermion masses one  needs
the Higgs bidoublet with the following quantum numbers:
$$\Phi=\left(
\begin{array}{ll}
\phi_1^0 & \phi_1^+ \nonumber \\
\phi_2^- & \phi_2^0
\end{array}
\right)=(\frac{1}{2}, \frac{1}{2}^*,0).
$$
This field  has to acquire nonzero v.e.v.,
saving however the electromagnetic invariance of vacuum:
$$ <\Phi>= \left(
\begin{array}{ll}
v_1 & 0 \\
0   &v_2
\end{array} \right)
$$
The existence of the abovementioned bidoublet is not enough to yield
the spontaneous symmetry breaking of the $SU(2)_L \otimes SU(2)_R$
gauge group \cite{3}.

 There may be an alternative choice of additional Higgs fields:
 a) Higgs doublets:
 \be
 \delta_L=\left[
 \begin{array}{l}
 \delta_L^+  \\
 \delta_L^0
 \end{array}
 \right]=(\frac{1}{2},0 ,1); \;
 \delta_R=\left[
 \begin{array}{l}
 \delta_R^+  \\
 \delta_R^0
 \end{array}
 \right]=(0, \frac{1}{2},1)
 \ee

They can generate heavy right-handed gauge bosons, but cannot
interact with fermions.

b) or Higgs triplets:
\be
\Delta_L= \left[
\begin{array}{l}
\Delta_L^{++} \\
\Delta_L^+ \\
\Delta_L^0
\end{array}
\right]=(1,0,2); \;\;
\Delta_R= \left[
\begin{array}{l}
\Delta_R^{++} \\
\Delta_R^+ \\
\Delta_R^0
\end{array}
\right]=(0,1,2)
\ee

They can produce large $M_{W_R}$ as well as Majorana masses for
neutrinos.

We remind that the  cancellation of all  quadratic divergencies
in the scalar sector can be produced only by the
compensation between the bosonic and fermionic loops.
As a result, in the case a) the fine tuning cannot
be implemented,
since no fermions with usual quantum numbers can
have Yukawa couplings to these fields.
In the case b) (with the triplet Majorana-Higgs  representation
$\Delta_R $)
  Majorana masses for right-handed neutrinos should be
  generated,
having the same order of magnitude as right-handed
gauge boson masses. Then one can expect the cancellation of the quadratic
divergences in the scalar sector. The presence of the
left-handed Higgs-Majoron fields $\Delta_L$
in general case is not necessary. But if these fields exist
-- for example in the case of the manifest LR symmetry
(see \cite{3}),
the vacuum expectation  of the left-handed Higgs-Majoron
should be  extremely small because of the upper bound
$\sim 1 eV$ on the left-handed Majorana neutrino masses
\cite{Hikasa}.

Thus the FT in the LR Model leads to the unambiguous determination
of the symmetry breaking sector of the theory.

\subsection*{3. Vacuum-Energy Cancellation Condition}

  It is well known that in the Left-Right symmetric model,
as well as in all non-supersymmetric models the vacuum energy
diverges like the forth power of the ultraviolet cutoff
scale. This divergence contains contributions with opposite
signs from bosons and fermions which can compensate each other.
As far as we treat the model as a low-energy
effective theory, the cutoff scales for bosons and fermions may
differ. In a SUSY underlying theory these scales are related
to heavy superparticle masses and their difference reflects
the soft SUSY breaking.  In this case it happens to be possible
to cancel vacuum-energy contributions by the
fine-tuning of the cutoff scales (fine-tuning for vacuum
energies \cite{13.6}).

We consider the universal cutoff $\Lambda_F$
for all fermions in
order to preserve the  horizontal symmetry and the universal
boson cutoff $\Lambda_B$ implying  a Grand
Unification at high energies.  Then the fine-tuning for vacuum
energies at one-loop level reads:

\vbox{
$$\frac{\Lambda_B^4}{\Lambda_F^4}
\equiv \alpha^2= \frac{4N_F}{2N_B+N_S}=
\left\{
\begin{array}{l}
96/50 \;\; \; \Delta_L  \; exist\\
96/44 \;\;\; without \; \Delta_L
\end{array}
\right.
$$  }

(Here $N_F=24$ is the number of fermionic degrees of freedom,
including quark colors
(left- and right-handed neutrinos are considered
here as one Dirac fermion), $N_B=15$ is the number of
the gauge bosons, $N_S=20; \; 14$ --number of
scalar fields, including Goldstone modes, with and without left
Higgs-Majoron fields $\Delta_L$ respectively.

Thus  we have $\alpha \approx 1.3856$ --if $\Delta_L$
exist or
$\alpha \approx 1.4771$ without $\Delta_L$.
 One can see that $\alpha$ in the LR-model is more
 close to unity than in the Standard Model or in
 the Two-Higgs Standard Model \cite{13.6,13.7,13.8}.

\subsection*{4. Quadratic Divergencies Cancellation and Right-Handed
Neutrino Masses.}

Let us remind that the scalar sector in the Weinberg-Salam
electroweak theory
contains quadratic divergences in the tadpole diagrams and in the scalar
particle self-energy.  In order to fit the finite masses
the cancellation for
 quadratic divergences, the fine-tuning
\cite{10.1} is required. This cancellation
holds only if fermion and boson loops are tuned due to specific
values of coupling constants. At the one
loop level the condition:
\be
(2M^2_W+M^2_Z+m^2_H)=\frac{4}{3}
\sum_{flavors, \; colors} m^2_f     \label{eq:Velt}
\ee
removes quadratic divergences both from the Higgs-field v.e.v. and the
Higgs boson self-energy if \underline{the universal momentum cutoff}
 is implemented for all the fields.

The original Veltman condition is however not stable under rescaling
since its renorm-derivative cannot vanish simultaneously for any
choice of the cutoff $\Lambda$ below the Plank scale \cite{10.13},
i. e. the required cancellation of quadratic divergences can be
provided only at a selected scale.

On the other hand, the usage of the universal scale for all bosons
and fermions roughly implies the existence of large symmetry involving
all the observable particles in one
multiplet and therefore is not well motivated
within the framework of effective theory.\\

We shall implement the Veltman's idea in the LR model for  the
case with different cutoff scales for bosons and fermions.  Let
us examine the  form of the Higgs-Yukawa and Higgs self-couping
lagrangian in the LR-model with different choice of particle
content.

  The general Higgs-Yukawa lagrangian for the
  bidublet field $\Phi$ and the three fermion generations
reads:
 $$L_{Yuk}=\overline{\psi}_L^i(F_{ij} \Phi +G_{ij} \tilde{\Phi})\psi_R^i+
 \overline{\psi}_R^i(F^*_{ij} \Phi^+ +G^*_{ij} \tilde{\Phi^+})\psi_L^i
 $$

  (Here $i,j=1,2,3$)

 Since in the
LR-model one has more scalar fields than in the
Standard Model, different ways of producing
fermion mass hierarchy are available.

 For our purposes we shall neglect the fermion masses of the first two
 generations, i.e. set $F_{ij}=G_{ij}=0$ for $i,j=1,2$.
In order to avoid problems, connected with the
  flavour-changing neutral currents, we shall set
the constant $G_{33} = 0$. Hence   only $F_{33} \equiv
 F\not=0$ and  the top-bottom mass difference is produced by
 the hierarchy of v.e.v's:  $m_t/m_b=v_1/v_2$.

 The Yukawa-type lagrangian for the Higgs-Majoron fields
may be written only for the lepton sector, because only
neutrinos can have Majorana masses. It has the form
(for right-handed fields):

 \be L_{MYu}\sim
 -\frac{h_M}{2}
 \overline{\omega}(i\tau_2 \Delta_R \frac{(1-\gamma_5)}{2}
 -\Delta_R^+i\tau_2\frac{(1+\gamma_5)}{2})\omega
 \ee
Here: $\omega \equiv \psi_R +C\overline{\psi}_R^T$;
and $\psi_R$ is the right-handed component
of Dirac  spinor  for the
weak isodoublet lepton fields;
$$\Delta_R=\left(
\begin{array}{lr}
\frac{\delta_R^+}{\sqrt{2}} & \delta_R^{++} \\
\delta_R^0 &   -\frac{\delta_R^+}{\sqrt{2}}
\end{array}            \right)
$$
For the theory with $\Delta_L$ fields one has the similar
lagrangian for the left Higgs-Majoron and left-handed
leptons, but we shall not write it down, since the
corresponding v.e.v. is very
small and does not influence on the heavy particle spectrum.

   The general form of Higgs potential of the model
  contains  15 (!) self-coupling vertices  \cite{3}
  and it is presented in the Appendix.
  For the simplicity we shall consider the
  situation, when  "non-diagonal" interactions (i.e. mixing
  fields $\Phi$ and $\Delta_{L,R}$ are suppressed.  This
  condition is one-loop  renorm-invariant for all
  ``non-diagonal" couplings except for $a_1$.  The latter one
  is assumed to be zero  only at the scale $\Lambda$, and
  its non-zero renorm-group flow is to be taken into account.

  Then the scalar potential is divided into two parts:
  the bidoublet potential and the triplet potential.
  The potential
   for the  $\Phi$  fields contains
  5 self-coupling vertices:
  $$V_{\Phi}\sim l_1 Tr^2(\Phi \Phi^+)+
  l_2[Tr^2(\tilde{\Phi} \Phi^+)+Tr^2(\tilde{\Phi^+}\Phi)]+
  l_3Tr(\tilde{\Phi}\Phi^+)Tr(\tilde{\Phi}^+\Phi)+$$
  $$+l_4Tr(\Phi \Phi^+)Tr[(\tilde{\Phi}\Phi^+) +
  (\tilde{\Phi}^+\Phi)]+a_1 \left[ Tr(\Phi \Phi^+) \right]
   \cdot Tr\left[
   (\Delta_L \Delta_L^+)
    + (\Delta_R \Delta_R^+) \right] $$
 Here $a_1(\Lambda)=0$.
The potential for the right Higgs-Majoron  fields
then reads (at the scale $\Lambda$):
$$V_{\Delta} \sim
\rho_1 tr^2(\Delta_R \Delta_R^+)
+\rho_2 tr(\Delta_R)^2tr(\Delta_R^+)^2 $$.

 Using this lagrangian one can derive the modified
 Veltman equations cancelling quadratic
 divergences for the scalar tadpoles of the
 theory.
 These equations for  the bidoublet fields $\Phi$  are --\\
  for the case without $\Delta_L$:
  \be \left\{
  \begin{array}{l}
  f_1\equiv 4F^2-\alpha[\frac{3}{2}g_L^2+\frac{3}{2}g_R^2
  +\frac{20}{3}l_1 +\frac{8}{3}l_3+ 4l_4\frac{v_2}{v_1}+2a_1] \\
  f_2\equiv 4F^2-\alpha[\frac{3}{2}g_L^2+\frac{3}{2}g_R^2
  +\frac{20}{3}l_1 +\frac{8}{3}l_3+ 4l_4\frac{v_1}{v_2}+2a_1]
  \end{array}
  \right.      \lbl{t}
  \ee
  and for the case with $\Delta_L$:
  \be \left\{
  \begin{array}{l}
  f_1\equiv 4F^2-\alpha[\frac{3}{2}g_L^2+\frac{3}{2}g_R^2
  +\frac{20}{3}l_1 +\frac{8}{3}l_3+ 4l_4\frac{v_2}{v_1}+4a_1] \\
  f_2\equiv 4F^2-\alpha[\frac{3}{2}g_L^2+\frac{3}{2}g_R^2
  +\frac{20}{3}l_1 +\frac{8}{3}l_3+ 4l_4\frac{v_1}{v_2}+4a_1]
  \end{array}
  \right.    \lbl{tl}
  \ee

  Here $a_1 \approx 0$.
  Hence if $m_t\not=m_b\,\Rightarrow l_4=0$.
  This condition is  renorm-invariant.
  The requirement $l_4=0$ follows also from the
  independent cancellation of quadratic divergences
  in the self-mass diagrams for  fields $\Phi$.


The modified Veltman condition for the right-handed
Higgs-Majoron fields $\Delta_R$ takes the form:
\be
2h_M^2= 3\alpha[2(2g^2+g'^2)+32\rho_1+16\rho_2].
\ee

Analyzing the modified Veltman conditions for the LR model,
one can find that in contrast to the Standard
Model, equations  \ceq{t}, \ceq{tl} do not yield
any bounds from below for
the $t-$quark mass, because not all $l_i$
constants must have positive values (for example
$l_3$ may be negative).  However such bounds
can be obtained for the right-handed Majorana
neutrinos.

>From the positive definiteness of the Higgs-Majoron potential
one can get that $\rho_1>0$, $\rho_1+\rho_2>0$. Then from the
modified Veltman condition it comes out that
\be 2h_M^2 \ge
6\alpha(2g^2+g'^2)
\ee
Thus right-handed Majorana
neutrinos must have masses of the same order of magnitude as
right-handed gauge bosons.
 Taking $M_R$ evaluations from different  experiments
 one obtains the corresponding lower bounds on $m_{\nu_R}$
 (see the Table 1).
\vspace{1cm}

\vbox{
\centerline{\bf Table 1}
 \vspace{0.1cm}
\vspace{0.5cm}
 \noindent
 \begin{tabular}{||l|c|c|c|c|c||} \hline
 Exp.data                  & $M_{W_R}$    & $m_{\nu_R}(\Delta_L)$  & $m_{\nu_R}$ & $m^3_{\nu_R}(\Delta_L)$ & $m^3_{\nu_R}$ \\ \hline
 $\Delta m_K+$             &   800         & 2.5                    & 2.3         & 1.44                    &  1.33       \\
 $+B\overline{B}_d+$       &   670          & 2.1                   & 1.9         & 1.21                    &  1.10        \\
$+b+\beta \beta$           &   740          & 2.31                  & 2.2         & 1.33                    &  1.26        \\
                           & (450 )        & (1.40 )               & (1.24 )     & (0.81 )                 &  (0.72 )        \\
man. LR                    & 1400           &  4.4                   & 4.0          & 2.52                    &  2.3         \\  \hline
 $\Delta m_K+$             & 500            &  1.56                 & 1.43        &                         &                \\
 $+B\overline{B}_d+$       & 500            &  1.56                 & 1.43        &                         &                \\
 $+\mu$                    & 740            &  2.30                 & 2.2         &                         &                 \\
 & (420 )                  &  (1.30 )       & (1.2 )                &                         &                 \\
man. LR                    & 1300           &  4.1                   & 3.8          &                         &                  \\ \hline
$m_{1R}< 10 MeV$,          & 720            &  2.2                  & 2.0         &                         &                 \\
Supernova                  & 16200          &  50                    & 46           &                         &                 \\
Dir. search                & 520; 610      &   1.6;1.9
& 1.59        &                          &                  \\
Rad.corr.                  & 439            &   1.36                 & 1.25
& 0.80                      & 0.74                 \\ \hline \end{tabular}
}

\vspace{2cm}

 The case of single heavy neutrino is represented in two
 central columns of the Table, the case of equal neutrino
 masses for all 3 generations - in two right columns. The left
 column of each pair corresponds to the theory with $\Delta_L$
 field ($\alpha=1.215$) while the right one - to the theory
 without $\Delta_L$  ($\alpha=1.276$). The left two columns
 are taken from \cite{01,3,520,439}.

 One should bear in mind that all the bounds on
 right-handed boson masses  are model dependent.
  All the data from experiments with
 strange mesons depend essentially on the right-handed
 Kobayashi-Maskawa matrix elements.
 For example the double $\beta$--decay
 data can be eliminated after fine-tuning in the corresponding
 leptonic mass matrix.  However, all these bounds
 are valid under rather reasonable assumptions and may
 be used for estimations of the right-handed neutrino masses.

       For the theory without left Higgs-Majoron ($\Delta_L$)
       the equality $g_L=g_R$ holds  only at the GUT scale.
       Then $\alpha_{2L}(m_Z)=0.0354$ implies $\alpha_{2R}=0.0265$.
       It is  taken into account in this table.

       The main result of the quadratic
       fine-tuning in the Higgs-Majoron sector
is that the absense of quadratic divergences leads to
rather heavy right-handed Majorana neutrinos.

\subsection*{5. RG-Stability of Fine-Tuning and Top-quark Mass.}

Let us apply the third of the Fine-Tuning principles (Weak
Fine-Tuning) --the RG-stability of
of quadratic divergencies cancellation-- in
order  to obtain estimations on the top-quark mass.
As it was shown in \cite{13.6} the RG stability
of the modified Veltman condition  in the
Standard Model
can be achieved
only due to different cutoff scales for bosons and fermions.
The abovementioned three Fine-Tuning principles lead
in the Standard Model to
the following top- and Higgs-mass predictions:
$m_t=175 \pm5$ GeV, $m_H=210 \pm10$ GeV.
In contrast ot the Standard Model, the LR model contains
more scalar degrees of freedom and its scalar self-couplings
may have different signs. However, the providing of
the RG-stability of leading divergencies
cancellation leads to a rather narrow range for the
$m_t$ values, which is compatible with the modern experimental
data.

Let us consider the case without $\Delta_L$.
The vacuum energy cancellation yields:
$\alpha=1.4771$.
 Using RG-equations (see Apppendix 1)
one can  get (assuming at the scale $\Lambda$ that
$g_L=g_R\equiv g$):

$$
Df_{\Phi}= 40 F^4 -F^2(64g_3^2 +36 g^2 +\frac{4}{3}g'^2) - \frac{\alpha}{3}
\left\{ 640 l_1^2  + 512 l_1 l_3 + 448l_3^2 +1304 l_2^2 +\right.
$$
\be \left.+240 l_1(F^2-\frac{3}{2}g^2)
+96l_3(F^2-\frac{3}{2}g^2) +28.5 g^4- 96 F^4 \right\} =0 \ee

Excluding $l_1$ by using $f_{\Phi}=0$:
$$l_1= \frac{3}{5\alpha}F^2 -\frac{9}{20}g^2 -\frac{2l_3}{5}$$
one has quadratic equation for $F^2$:
$$c_1F^4+c_2F^2 +c_3=0;$$
here:
$$c_1=32\alpha-8-\frac{384}{5\alpha}\approx -12.727$$
$$c_2=-64g_3^2-36g^2-\frac{4}{3}g'^2+ \frac{576}{5}g^2 +72g^2
+36\alpha g^2$$
$$
c_3= -\alpha(\frac{144}{5}l_3+96l_2+106.7g^4)$$

It can be easily checked up that this equation can have positive solutions
only for such values of gauge couplings which they
have at energies much more than $100 \, GeV$.
Let us assume that at the scale $\Lambda$ the left-handed
and the right-handed couplings are equal: $g_L=g_R$
(in the absense of $\Delta_L$ fields they have different RG-flows).
Then the solutions of the above equations result in rather
narrow range of possible values for $m_t$ for different
 $\Lambda$ (see  two left columns of the Table 2).

In the case with $\Delta_L$ the vacuum energy cancellation yields
$\alpha \approx 1.3856$ and the RG-flow for $g_L$
is the same as for $g_R$ (see Appendix 1)
Then the equation for $F^2$
$$c_1F^4+c_2F^2 +c_3=0;$$
has the coefficients:
$$c_1=32\alpha-8-\frac{384}{5\alpha}\approx -19.085$$
$$c_2=-64g_3^2-36g^2-\frac{4}{3}g'^2+ \frac{576}{5}g^2 +72g^2
+36\alpha g^2$$
$$
c_3= -\alpha(\frac{144}{5}l_3+96l_2+110.2g^4)$$

Possible solutions are displayed in two right
columns of the Table. They lead
to more strict bounds for possible $\Lambda$ scale and
allowed values for $m_t$.

\vbox{
\begin{center}
{\bf Table 2.  Masses of the $t$-quark in the LR Model.}

~~~~~~~~Case  without  $\Delta_L$. ~~~~~~~~~~Case with $\Delta_L$.

\begin{tabular}{||l|c|c|c|c||} \hline \hline
$\Lambda \, GeV$ &  $``m_t(\Lambda)"$ &$m_t(100 \, GeV)$ &  $``m_t(\Lambda)"$ &$m_t(100 \, GeV)$ \\ \hline
$10^{15}$        &   107--287  &166--202  &  110--225  & 167--197           \\ \hline
$10^{14}$        &   110--287  & 167--205 &   113--224 &169--199          \\ \hline
$10^{13}$        &  113--287   & 168--208 & 116--223   & 170--201          \\ \hline
$10^{12}$        &  116--285   & 169--211  & 121--221   & 172--203          \\ \hline
$10^{11}$        &  120--283   & 171--214  & 126--217   & 175--205          \\ \hline
$10^{10}$        &  125--279   & 173--218  & 133--212   & 178--207          \\ \hline
$10^{9}$         &  132--273   & 177--222  & 144--202   & 184--207          \\ \hline
$10^{8}$         &  141--264   & 181--225  & ---        & ---          \\ \hline
$10^{7}$         &  155--247   & 189--227  & ---        & ---            \\ \hline
$10^{6}$         &   ---       & ---       &  ---       & ---                 \\ \hline
\end{tabular}
\end{center} }

The denotation $``m_t(\Lambda)"$  means $g_t(\Lambda) \cdot 175$ GeV.

To predict the top-quark mass  $m_t(100\, GeV)$
we need to use the RG flow:
$$F^2(\mu)=\left(\frac{g_3^2(\mu)}{g_3^2(\Lambda)}\right)^{8/7}
\left(\frac{g^2(\mu)}{g^2(\Lambda)}\right)^{3/4}
\left(\frac{g'^2(\mu)}{g'^2(\Lambda)}\right)^{17/84}\cdot $$
$$\cdot \left\{
1+\frac{5F^2(\Lambda)}{8\pi^2}\int_{\mu}^{\Lambda}dt
\left(\frac{g_3^2(t)}{g_3^2(\Lambda)}\right)^{8/7}
\left(\frac{g^2(t)}{g^2(\Lambda)}\right)^{3/4}
\left(\frac{g'^2(t)}{g'^2(\Lambda)}\right)^{17/84}\right\}^{-1}.$$

In the Table 2  the largest
and the smallest values of $m_t$ are shown. They
correspond to the choice $l_3=l_2=0$, while nonzero values
of these self-couplings push two possible values of the top
mass inside the interval.
One can see that the above equation contains
restrictions on the maximal possible values of the $l_3$ and
$l_2$.  For the gauge couplings the experimental input was
taken as follows \cite{Hikasa}:  $$\alpha_3(M_Z) \equiv
\frac{g_3^2}{4 \pi}= 0.118 \pm 0.007  $$ $$\alpha_{em} \equiv
\sin^2  \theta_W \frac{g_L^2(M_Z)}{4\pi}= (127.8\pm0.1)^{-1} $$
$$\sin^2 \theta_W(M_Z) = 0.2333 \pm 0.0002 $$

  One can notice that the experimental value $m_t \approx 175$ GeV
is compatible with the range given in the Table 2.
The presented estimations are rather sensitive to the input
values of the gauge couplings, especially to $\alpha_3$.

  \subsection*{6. Conclusion}

We have shown that in the Left-Right symmetric Model
as well as in the  Standard Model with one and two Higgs
doublets \cite{13.6,13.7} the selection rule
based on the vacuum fine-tuning can be implemented for the parameters
of $t$-quark and Higgs-boson
potentials. This rule requires the existence
of the right-handed Majorana neutrinos and yields
lower bounds on their masses.
The FT conditions for the $t$--quark parameters lead to
predictions of the $t$--mass in a good agreement
with the experimental value.
The approximate
RG invariance, which is used for these predictions,
 is an important property of the fine-tuning conditions
which otherwise do not acquire the universal meaning.
We notice that in the LR model it could be very interesting
to analyze a more general form of the scalar potential,
using the available experimental bounds
on its constants.\\

This work is partially
supported by the RFBR Grant No.95-02-05346a
and by the GRACENAS grant No.96-6.3-13.

  \subsection*{APPENDIX 1. Scalar Potential in the
  Left-Right Symmetric Model.}

This is the general form  of the scalar potential
in the LR model with bidublet and left and right
Higgs-Majoron fields:
$$
 V= -\mu^2_1 \left[Tr(\Phi^+ \Phi)  \right]
    -\mu^2_2 \left[Tr(\tilde{\Phi} \Phi^+)
    +Tr(\tilde{\Phi}^+ \Phi \right)] - $$
 $$
    -\mu_3^2 \left[ Tr(\Delta_L \Delta_L^+)
    + Tr(\Delta_R \Delta_R^+)\right]+
    $$
    $$
    + l_1 Tr^2(\Phi \Phi^+)+
  l_2[Tr^2(\tilde{\Phi} \Phi^+)+Tr^2(\tilde{\Phi^+}\Phi)]+
  $$
  $$
  +l_3Tr(\tilde{\Phi}\Phi^+)Tr(\tilde{\Phi^+}\Phi)
  +l_4Tr(\Phi \Phi^+)[Tr(\tilde{\Phi}\Phi^+) +
  Tr(\tilde{\Phi^+}\Phi)]+
 $$
 $$
 +\rho_1  \left[ Tr^2(\Delta_L \Delta_L^+)
    + Tr^2(\Delta_R \Delta_R^+)\right] + $$
    $$+\rho_2
    \left[ Tr(\Delta_L \Delta_L) Tr(\Delta_L^+ \Delta_L^+)
    + Tr(\Delta_R \Delta_R) Tr(\Delta_R^+ \Delta_R^+)\right] +
    $$
    $$+
  \rho_3 \left[  Tr(\Delta_L \Delta_L^+) Tr(\Delta_R \Delta_R^+)
 \right]+ $$
 $$
 +\rho_4 \left[Tr (\Delta_L \Delta_L) Tr(\Delta_R^+ \Delta_R^+)
 +Tr (\Delta_L^+ \Delta_L^+) Tr(\Delta_R \Delta_R)\right]+
 $$
 $$+ a_1 \left[ Tr(\Phi \Phi^+) \right]
   \cdot \left[
   Tr(\Delta_L \Delta_L^+)
    + Tr(\Delta_R \Delta_R^+) \right]+ $$
  $$
  +  a_2\left\{ \left[ Tr(\Phi \tilde{\Phi}^+) \right] \cdot
   Tr(\Delta_R \Delta_R^+) +
   \left[Tr(\Phi^+ \tilde{\Phi}) \right]
   Tr(\Delta_L \Delta_L^+)\right\}+
  $$
  $$
  +  a_2^* \left\{ \left[ Tr(\Phi^+ \tilde{\Phi}) \right] \cdot
   Tr(\Delta_R \Delta_R^+) +
   \left[Tr(\tilde{\Phi}^+ \Phi) \right]
   Tr(\Delta_L \Delta_L^+)\right\}+
  $$
  $$
  +a_3\left[Tr(\Phi \Phi^+ \Delta_L \Delta_L^+)
  +Tr(\Phi^+ \Phi \Delta_R \Delta_R^+)
  \right] +
  $$
  $$
  + \beta_1 \left[Tr (\Phi \Delta_R \Phi^+ \Delta_L^+)+
  Tr (\Phi^+ \Delta_L \Phi \Delta_R^+)
  \right] +
  $$
  $$
  +  \beta_2 \left[Tr (\tilde{\Phi} \Delta_R \Phi^+ \Delta_L^+)+
  Tr (\tilde{\Phi}^+ \Delta_L \Phi \Delta_R^+)
  \right] +
  $$
  $$
  +  \beta_3 \left[Tr (\Phi \Delta_R \tilde{\Phi}^+ \Delta_L^+)+
  Tr (\Phi^+ \Delta_L \tilde{\Phi} \Delta_R^+)
  \right] .
  $$

  \subsection*{APPENDIX 2. Renorm-Group Equations for the
  Left-Right Symmetric Model.}

Case without left Higgs-Majoron fields:

$$D g'=\frac{10}{3}g'^3$$
$$Dg_L=-3g_L^3$$
$$Dg_R=-\frac{7}{3}g_R^3$$

$$Dg_3=-7g_3^3$$

$$DF= F \left(5F^2 -8g_3^2 - \frac{9}{4}(g_L^2+g_R^2) - \frac{1}{6}g'^2)
\right)$$

$$ D l_1= 32 l_1^2 +   16 l_1 \cdot l_3
+16 l_3^2 +64 l_2^2 $$
$$-3 l_1(3g_L^2+3g_R^2) +
12 l_1 F^2+\frac{9}{8}(g_L^4 +\frac23 g_L^2 g_R^2+g_R^4)
-6F^4 $$

$$D l_3= 8 l_1 \cdot l_3+
16(l_1+l_3)\cdot l_3 +16 l_2^2 -3 l_3(3g_L^2+3g_R^2)+$$
$$+12 l_3 \cdot F^2 +12F^4+6g_L^2 g_R^2 $$

$$
D a_1/_{a_1=0}=\frac34g^4
$$

For the case with left Higgs-Majoron fields

$$Dg_{L,R}=-\frac{7}{3}g_{L,R}^3; \;\; Dg'=4g'^3$$
(So the equality $g_L=g_R$ holds for any energy.)
All other equations are the same.

\end{document}